\documentclass[aps,reprint,superscriptaddress,amsmath,amssymb]{revtex4-2}

\usepackage{graphicx}
\graphicspath{{figures/}}
\usepackage{listings}
\usepackage{longtable}
\usepackage{color}
\usepackage{xcolor}
\usepackage[colorlinks,allcolors=blue]{hyperref}
\usepackage{algpseudocode}
\usepackage{algorithm}
\newcommand\ket[1]{\ensuremath{|#1\rangle}}
\newcommand\bra[1]{\ensuremath{\langle#1|}}
\newcommand{\relu}{\operatorname{ReLU}}
\newcommand{\hz}{\mathrm{Hz}}

\def\<{\langle}
\def\>{\rangle}

\begin{document}
\title{Integrating Quantum Processor Device and Control Optimization\\ in a Gradient-based Framework}

\author{Xiaotong Ni}
\affiliation{Alibaba Quantum Laboratory, Alibaba Group, Hangzhou, Zhejiang 311121, P.R.China}

\author{Hui-Hai Zhao}
\affiliation{Alibaba Quantum Laboratory, Alibaba Group, Beijing 100102, P.R.China}

\author{Lei Wang} 
\affiliation{Institute of Physics, Chinese Academy of Science, Beijing, 100190, P.R.China}
\affiliation{Songshan Lake Materials Laboratory, Dongguan, Guangdong 523808, P.R.China}

\author{Feng Wu}
\affiliation{Alibaba Quantum Laboratory, Alibaba Group, Hangzhou, Zhejiang 311121, P.R.China}

\author{Jianxin Chen}
\affiliation{Alibaba Quantum Laboratory, Alibaba Group USA, Bellevue, Washington 98004, USA}

\begin{abstract}
In a quantum processor, the device design and external controls together contribute to the quality of the target quantum operations.
As we continuously seek better alternative qubit platforms, we explore the increasingly large device and control design space. Thus, optimization becomes more and more challenging.
In this work, we demonstrate that the figure of merit reflecting a design goal can be made differentiable with respect to the device and control parameters. In addition, we can compute the gradient of the design objective efficiently in a similar manner to the back-propagation algorithm~\cite{rumelhart_learning_1986} and then utilize the gradient to optimize the device and the control parameters jointly and efficiently. This extends the scope of the quantum optimal control to superconducting device design. We also demonstrate the viability of gradient-based joint optimization over the device and control parameters through a few examples.

\end{abstract}

\maketitle

Quantum mechanics provides an entirely new way to think about information processing. Information is stored in quantum systems and the dynamics are described by the Schr\"{o}dinger equation or the master equation.
To achieve desired operations, we need to design systems with suitable Hamiltonians and dissipation as well as time-dependent external controls.
Concrete examples of such quantum information processing tasks include quantum computing, quantum simulation, quantum error correction, and quantum metrology.
At this level of abstraction, it is clear that we are dealing with optimization problems that consist of both Hamiltonian design and control.
Traditionally, the optimization of the design~\cite{menke_automated_2021,liu2021quantum} and the optimization of the control~\cite{khaneja_optimal_2005,machnes_tunable_2018} are usually studied separately.
In this work, we treat them as equal components in a single optimization problem and use gradient information to speed up optimization.

We consider superconducting circuits, which are one of the most promising hardware platforms due to their device design versatility and fabrication scalability.
While the design versatility allows for many qubit and coupling types~\cite{ yamamoto_demonstration_2003, chiorescu_coherent_2003, Izmalkov_2004, Plourde_2004, koch_charge-insensitive_2007, majer_coupling_2007, barends_superconducting_2014, manucharyan_fluxonium_2009, yan_tunable_2018, ofek_extending_2016}, it also makes the quest of finding the best design harder due to the larger parameter space.
As mentioned earlier, the optimization of the device design is further complicated by the fact that we need to consider the control schemes simultaneously, which has been noted in previous work~\cite{chu_coupler-assisted_2021,goerz_charting_2017}.
For example, complex control schemes are needed for bosonic code qubits, and often numerical optimization is required to obtain the best gate performances~\cite{heeres_implementing_2017, hu_quantum_2019, wang2021automated}.
In scenarios where we cannot estimate the control performances with analytical formulas, the optimization problem of the qubit design inevitably becomes a joint optimization problem of design and control with an even larger parameter space.
A common strategy to handle large optimization problems is using efficiently computed gradients to help us traverse the optimization landscape.
One milestone in this direction is the \textbf{GR}adient \textbf{A}scent \textbf{P}ulse \textbf{E}ngineering (GRAPE) algorithm~\cite{khaneja_optimal_2005}.
In this work, we showed that the figure of merit of a target goal such as the gate fidelity can be made differentiable with respect to the parameters from the device and the control together.
The gradients can be computed in a single computation similar to the GRAPE algorithm and the back-propagation algorithm~\cite{rumelhart_learning_1986}.
More accurately, the ratio of the time needed to compute the figure of merit and its gradient is independent of the number of control parameters.
This makes our method feasible not only for optimizing complex pulse shapes but also for optimizing large processor device designs, and more interestingly, jointly optimizing the design and control. 

\begin{figure}
    \includegraphics[width=0.9\linewidth]{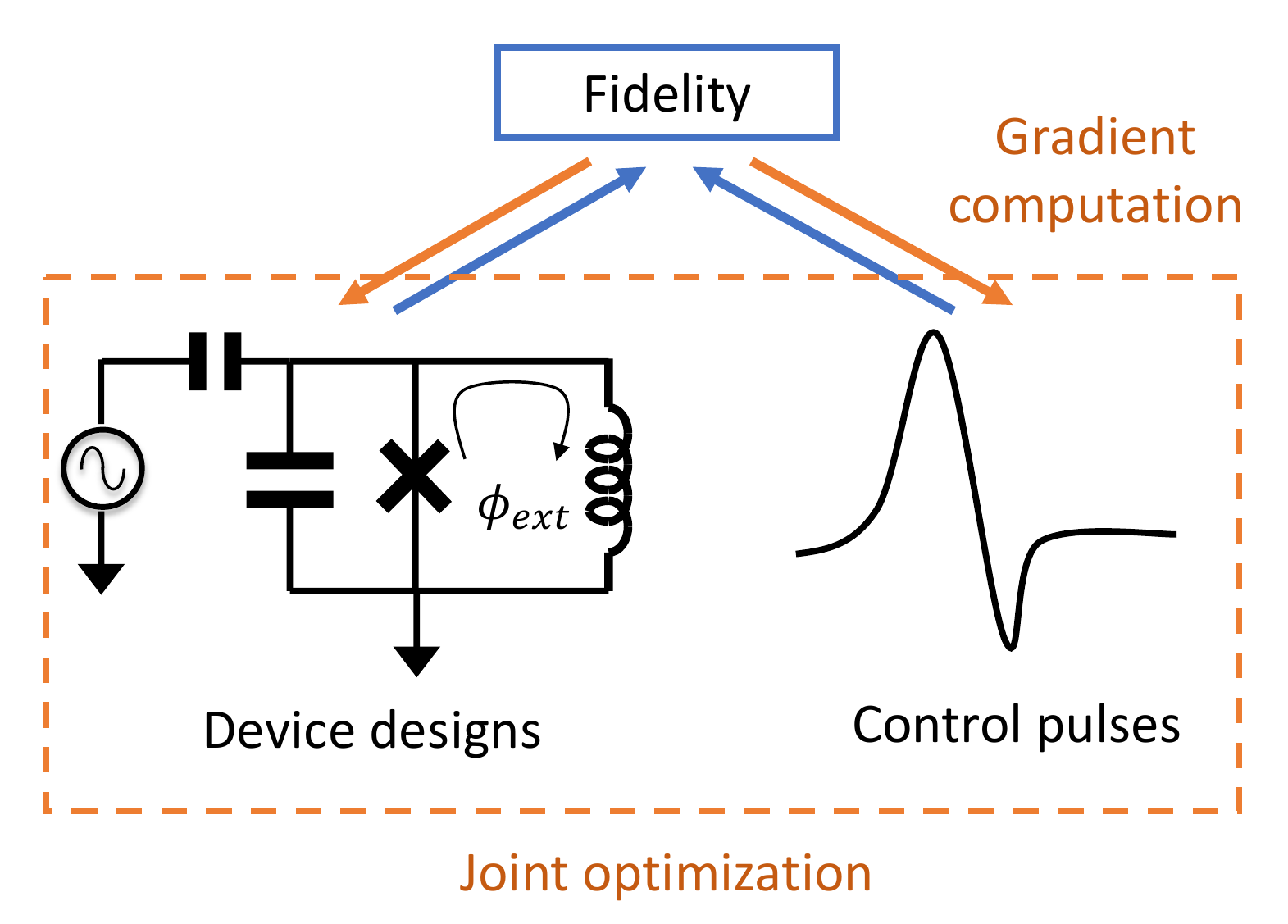}
    \caption{\label{fig_joint_optimization_concept} Illustration of joint optimization in circuit quantum electrodynamics systems. Both device designs and control pulses contain many parameters, and together they determine the fidelity of the quantum operations. In this work, we consider the problem of optimizing them together.}
\end{figure}

\label{sec_method}

\textit{Integrating design and control aspects.} 
In this work, we mainly focus on the two stages of quantum processor design that involve finding optimal values of the device parameters in the system Hamiltonian and deriving optimal control pulses to maximize the performance of the target quantum processor.
When the device parameters are determined and we only need to optimize over the control parameters for better performance or robustness, this is precisely what quantum optimal control and robust control do. In contrast, we may sometimes need to optimize the spectral or other properties of the device Hamiltonian itself, e.g., for determining tunable $ZZ$-interaction on--off ratios~\cite{dicarlo_demonstration_2009, stehlik_tunable_2021} or for designing 4-local interaction couplers~\cite{menke_automated_2021}. 

Now we introduce our framework that unifies the above two scenarios, optimizing a system's performance over the control parameters and the device parameters jointly
as illustrated in \autoref{fig_joint_optimization_concept}. 
The Hamiltonian of a quantum system with control fields takes the following form:
\begin{equation}
    H\left(\vec{h}, \vec{c},t\right)=
    H_\mathrm{dev}\left(\vec{h}\right)+\sum_{k=1}^{N_c}f_{k}\left(\vec{c},t\right)C_{k}\left(\vec{h}\right),
    \label{eq_general_hamiltonian}
\end{equation}
where $H_\mathrm{dev}$, $C_{k}$, and $f_{k}$ are the device Hamiltonian, control operators, and control fields, respectively, and $\vec{h}$ and $\vec{c}$ denote the device and control parameters, respectively. The goal is to optimize $\vec{h}$ and $\vec{c}$ for certain objectives, such as
implementing a specific unitary operator within a given time $\tau$.
This extended setup formulates a device-control codesign problem.
In~\autoref{section_opt_separately_simultaneous}, we explain why we expect that optimizing $\vec{h}$ and $\vec{c}$ simultaneously will be more efficient compared to the case where we artificially break the optimization problem into two and optimize $\vec{h}$ and $\vec{c}$ alternatively.
Compared to the optimal control problem where $\vec{h}$ is fixed, the extended setup will open more possibilities. For example, we can optimize the device parameters $\vec{h}$ for a gate implementation that is robust against control fluctuations and deviations. We will later showcase this by using our method to obtain a more robust iSWAP gate between capacitively coupled fluxonium qubits.
In comparison, with conventional quantum control, we only optimize over control parameters for robust schemes. 

\textit{Efficient gradient computation for both device and control parameters.}
Here, we discuss an efficient method for computing the gradient of both the device and control parameters.
More explanation about why we consider this method to be efficient can be found in~\autoref{subsect_autodiff}.
For a superconducting processor using qubits as its basic components, the total Hilbert space can be decomposed correspondingly as $\mathcal{H} = \bigotimes_i \mathcal{H}_i$, where each $\mathcal{H}_i$ is typically an infinite-dimensional Hilbert space.
The time-independent Hamiltonian $H_\mathrm{dev}$ in~\autoref{eq_general_hamiltonian} can be typically described by the Hamiltonian
\begin{equation}
    H_\mathrm{dev}(\vec{h}) = \sum_i H_{i}(\vec{h}) + \sum_{i,j} g_{ij}(\vec{h}) S_i(\vec{h}) S_j(\vec{h}), 
\end{equation}
where $H_i$ and $S_{i}$ are single component Hamiltonian and coupling operators defined on $\mathcal{H}_i$, respectively.
The concrete form of $S_{i}$ depends on the type of coupling between qubits.
For notational simplicity, we assume that there is a single type of coupling, but our computation method naturally generalizes to the situation where there are multiple types.
We will also assume that there is a single control operator $C_i(\vec{h})$ (see~\autoref{eq_general_hamiltonian}) for each subsystem $\mathcal{H}_i$.

Even though each component $\mathcal{H}_i$ has infinite dimensions, typically only the lowest few energy eigenstates will participate in the computation.
This is indeed the case for the examples presented in this work.
Therefore, an efficient method for performing numerical simulations is to first truncate the total Hilbert space into this relevant subspace $\bigotimes_i \mathcal{H}_i'$, where $\mathcal{H}_i'$ is the low energy subspace under the above assumption.
This can be done by diagonalizing $H_{i}(\vec{h})$.
We can then use the operators from projecting $H_i$, $S_i$, and $C_i$ into the subspace $\mathcal{H}_i'$ as an approximation and obtain an efficient finite dimensional Hamiltonian $H'$ for numerical computation.
Both the Hamiltonian diagonalization and the Hilbert space truncation can be made differentiable. Please refer to~\autoref{subsect_autodiff} for technical details. To the best of our knowledge, this is the first attempt to extend the GRAPE algorithm~\cite{khaneja_optimal_2005} to superconducting circuit device parameters.

The evolved unitary operator at time $\tau$ can be then derived by solving the ordinary differential equation (ODE)
\begin{equation}
    i\hbar \frac{d}{dt} U(t)= H(t)U(t).
\end{equation}
After we compute $U(\tau)$, we can use objective functions such as the average gate fidelity~\cite{nielsen_quantum_2000} to compute how close $U(\tau)$ and the desired operator $U_{\mathrm{target}}$ are.
In general, we need to add more terms to the objective functions to ensure the devices are practical.
For now, we assume that it is easy to perform the reverse-mode differentiation for the computation steps to compute the objective function $O(\vec{h},\vec{c})$ from $U(\tau)$.
To make the whole design workflow differentiable, we still need to perform reverse-mode differentiation through the ODE solver. The adjoint sensitivity method~\cite{pontryagin1962mathematical} makes this possible. This method computes gradients by solving a second, augmented ODE backwards in time, independent of the parameter size.  With the gradients computed in reverse mode, we can now optimize over the device and control parameters jointly and efficiently for an objective $O(\vec{h},\vec{c})$. 

The proposed framework may also be used to optimize for solutions robustly against control noise. Robust optimization over control parameters has been extensively studied ~\cite{zhang_robust_2014,chen_sampling-based_2014,kabytayev_robustness_2014}. Here, we take a sample-based approach~\cite{zhang_robust_2014,chen_sampling-based_2014} and define
\begin{equation}
    O_\mathrm{avg}(\vec{h},\vec{c}) = \frac{1}{N} \sum_{i=1}^N O(\vec{h},\vec{c} + \overrightarrow{\Delta c_i}),
    \label{eq_robust_objective_by_average}
\end{equation}
where $O(\vec{h},\vec{c})$ is the original design goal, e.g., the fidelity or another figure of merit, $N$ is the number of samples, and $\overrightarrow{\Delta c_i}$ is the deviation of $\vec{c}$ caused by noise in the $i$-th sample. Optimizing $O_\mathrm{avg}$ will make parameters $(\vec{h},\vec{c})$ robust against the selected noise.
Since we can compute the gradient of $O(\vec{h},\vec{c})$, we can compute the gradient of $O_\mathrm{avg}(\vec{h},\vec{c})$ by summing over the gradients of $O(\vec{h},\vec{c} + \overrightarrow{\Delta c_i})$.
In this way, we can also use the gradient information to speed up the robustness optimization.

As mentioned above, there are scenarios where we want to optimize the properties of $H_\mathrm{dev}(\vec{h})$ alone. 
In~\autoref{subsect_adiabatic_cphase}, we will show how we can choose an objective function for a large $ZZ$-interaction on--off ratio between two transmon qubits and the gradient computation method.

\label{subsection_iswap_time_evo_opt}

\begin{figure}
    \includegraphics[width=0.7\linewidth]{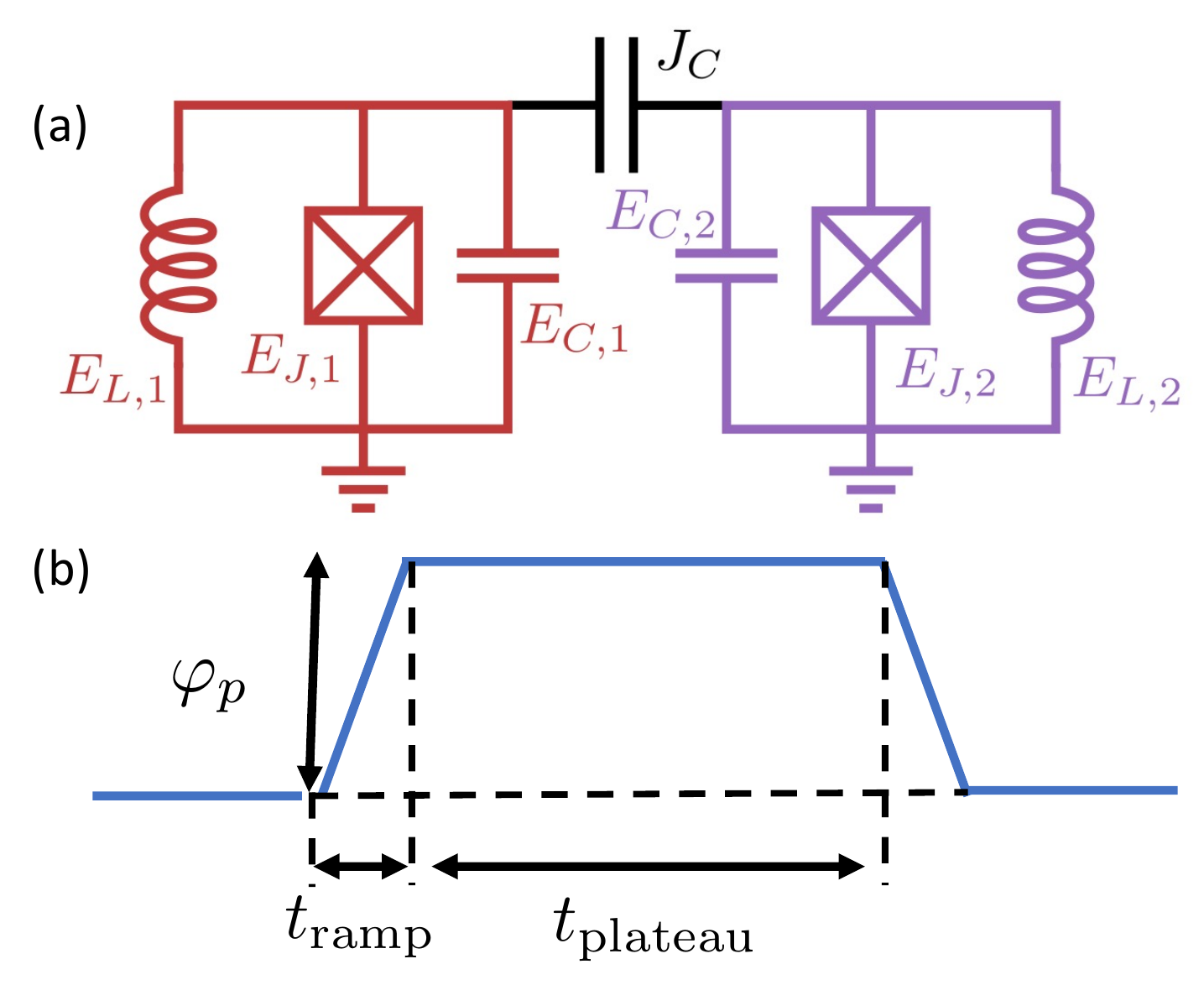}
    \caption{\label{fig_ctrl_device_params} Device and control parameters optimized for the \textrm{iSWAP} gate. (a) Circuit diagram of two fluxonium qubits coupled by a capacitor. Each circuit component corresponds to a term in the Hamiltonian, and we label the components with their energy parameters. The capacitive coupling strength $J_C$ is determined by the capacitance  of each qubit and between these two qubits. (b) Time-dependent part of the control pulse shape. It corresponds to the second term on the right-hand side in~\autoref{eq_rectangular_pulse}.}
\end{figure}

\textit{\textrm{iSWAP} gate with fluxonium qubits} We will demonstrate our method for an \textrm{iSWAP} between two capacitively coupled fluxonium qubits, 
the diagram of which is depicted in \autoref{fig_ctrl_device_params}(a).
The Hamiltonian of the system is
\begin{equation}
    H(t) = H_{f,1}+H_{f,2}(t) + J_C n_1 n_2, 
    \label{eq_time_dependent_capacitive_coupling}
\end{equation}
where $J_C$ is the coupling strength between the two qubits, and $H_{f,i}$ is the Hamiltonian of the $i$-th fluxonium qubit~\cite{manucharyan_fluxonium_2009}, defined as follows:
\begin{equation}
    H_{f,i} = 4E_{C,i} n_i^2 + \frac{1}{2} E_{L,i} (\varphi_i +\varphi_{\mathrm{ext},i})^2
    -E_{J,i}\cos \left( \varphi_i \right),
\end{equation}
where $E_C$, $E_J$, and $E_L$ are the charging energy, the Josephson energy, and the inductive energy, respectively, $\varphi_i$ is the phase operator, and $n_i$ is the conjugate charge operator.
We set the external flux of the second qubit to be a time-dependent control field initially at $\pi$, while for the first qubit, the external flux is fixed to be $\pi$, that is,  $\varphi_{\mathrm{ext},1}=\pi$ and $\varphi_{\mathrm{ext},2}(0) = \pi$.
The waveform of $\varphi_{\mathrm{ext},2}(t)$ is chosen to be a trapezoid pulse as shown in \autoref{fig_ctrl_device_params}(b), which is given by
\begin{multline}
\varphi_{\mathrm{ext},2}(t) = \pi+ \\
\varphi_{p} \min\{\mathrm{ReLU}(\frac{t}{t_{\mathrm{ramp}}}),1,
\mathrm{ReLU}(\frac{2t_{\mathrm{ramp}}+t_{\mathrm{plateau}}-t}{t_{\mathrm{ramp}}})\},
\label{eq_rectangular_pulse} 
\end{multline}
where $\varphi_p$ and $t_{\mathrm{plateau}}$ are the external flux bias and holding time of the plateau of the waveform, $t_\mathrm{ramp}$ is the time for the rising and falling edges, and $\mathrm{ReLU}(x) = \max (x,0)$ is a commonly used activation function in neural networks.
We call $E_C$, $E_J$, $E_L$, and $J_C$ the device parameters, and the parameters in $\varphi_{\mathrm{ext},2}(t)$ are the control parameters \footnote{Though $\varphi_{\mathrm{ext},2}(t)$ was part of the device Hamiltonian, it comes from the external flux control, so the parameters therein are more appropriately regarded as control parameters.}. These are all parameters involved in the optimization.

Some details of the construction of the finite dimensional Hamiltonian are provided in what follows.
We first write the Hamiltonian of the single-qubit fluxonium in the discretized phase basis over a finite range of phase values. To obtain sufficiently accurate low-energy eigenstates for the optimization task, we select the range to be $\varphi \in [-5\pi, 5\pi]$ and the number of basis points to be $n_\mathrm{basis}=400$. 

We select the average fidelity to be the objective function.
The leakage is small for the iSWAP gate scheme, and the optimized results have a leakage smaller than $10^{-4}$.
Therefore, it is a good approximation to compute the fidelity based on the truncated $4\times 4$ unitary matrix.
However, there are usually single qubit phases accumulated during the implementation of the gate, and we can compensate for these phases at a very low cost~\cite{mckay_efficient_2017}.
Therefore, we will use a modified average fidelity $F_m(U)$ with phase compensation included (see~\autoref{subsection_fidelity_phase_compensation}).
We include other terms in the total objective function.
One is a penalty $P_{\mathrm{decoh}}$, which roughly estimates the infidelity caused by the decoherence of $T_1$ and $T_2$.
Another is $P_\mathrm{fDiff}$, which discourages the $0 \leftrightarrow 1$ frequencies of the two fluxonium qubits from being too close.
This is useful for having addressable single qubit gates and it mitigates $ZZ$-crosstalk when both fluxonium qubits are at their sweet spots.
The last one is $P_\mathrm{fm}$, which forces the fluxonium parameters to be realistic for fabrication.
The exact forms of these terms will be given in~\autoref{section_iswap_opt_details}.

In total, we set the objective function for minimization to be
\begin{equation}
    O = \ln (1-F_m(U)+P_{\mathrm{decoh}}+P_\mathrm{fDiff}+P_\mathrm{fm}).
    \label{eq_iswap_objective}
\end{equation}

\begin{figure}
    \includegraphics[width=0.9\linewidth]{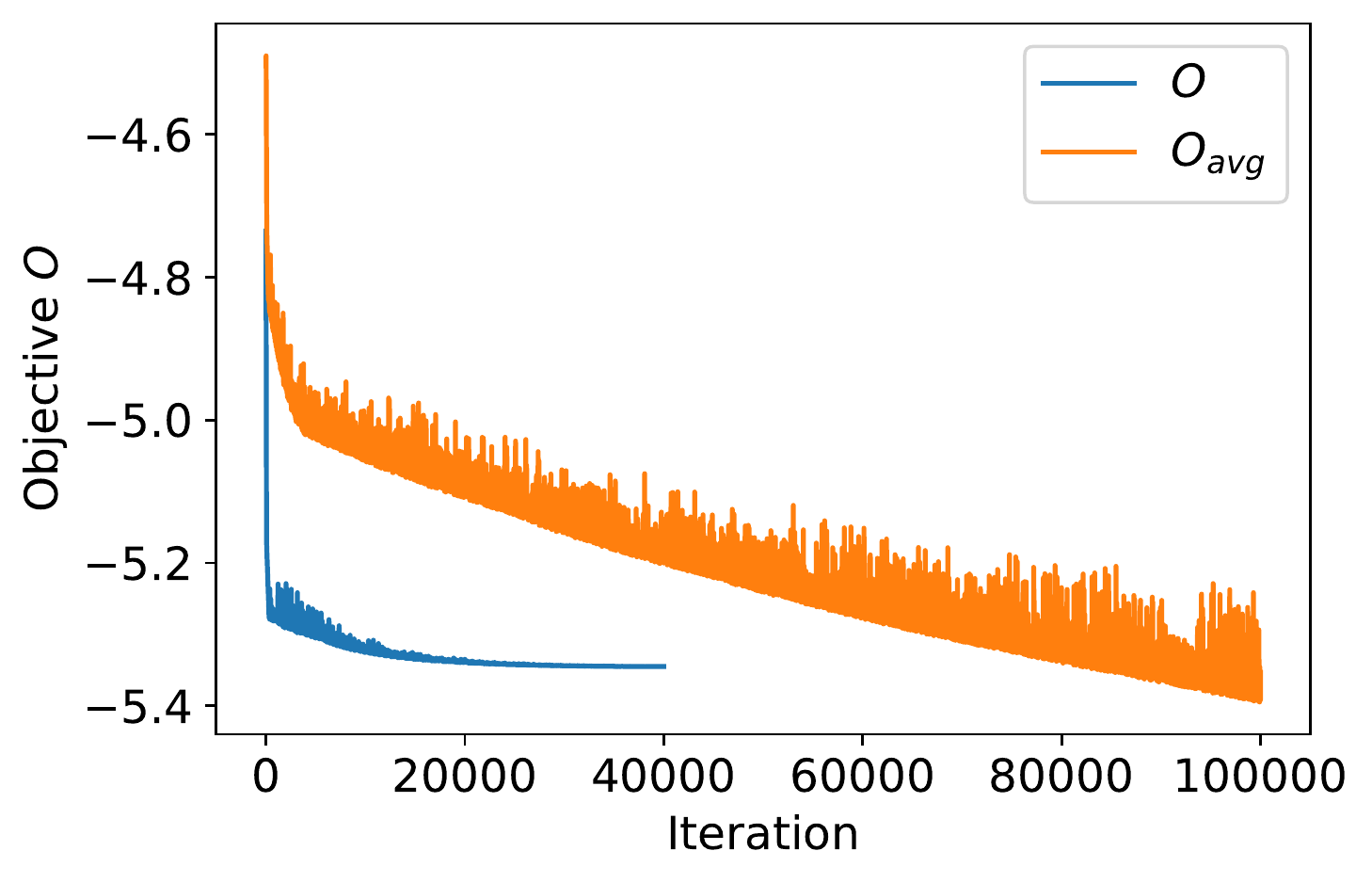}
    \caption{\label{fig_iswap_loss}iSWAP gate optimization loss during training.
    The blue curve corresponds to the optimization of $O$ in~\autoref{eq_iswap_objective} and the orange curve corresponds to $O_{\mathrm{avg}}$ in~\autoref{eq_robust_objective_by_average}.}
\end{figure}

\begin{figure}
    \includegraphics[width=0.9\linewidth]{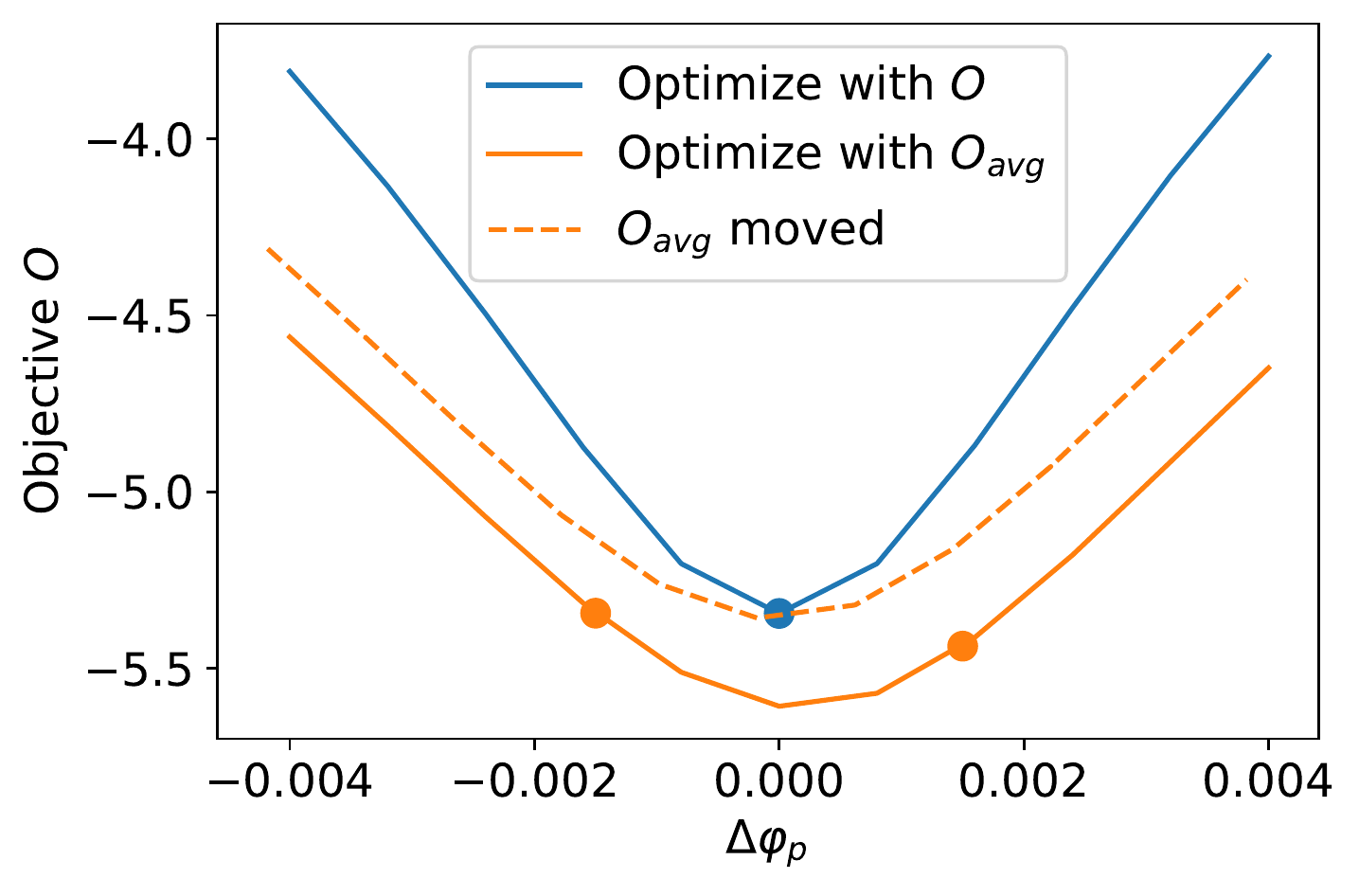}
    \caption{\label{fig_opt_robust}Robustness of the iSWAP gate with respect to the control parameter $\varphi_p$. For different $\Delta \varphi_p$, we compute $O(\operatorname{add}(\vec{p},\varphi_p,\Delta \varphi))$. The blue and orange dots mark the $\Delta \varphi_p$ values used to perform the optimization. The orange dotted curve is a translation of the orange curve. The curvature of the orange dotted curve is smaller than that of the blue curve. Therefore, the parameters obtained from the robustness optimization are more robust.}
\end{figure}

Another desired feature of the \textrm{iSWAP} gate scheme is robustness against control errors.
Here, we demonstrate how we can improve the robustness of the gate fidelity $F_m(U)$ with respect to the amplitude parameter $\varphi_p$ in~\autoref{eq_rectangular_pulse}.
To do this, we set $\vec{\Delta c_{1,2}} = (\pm \Delta \varphi_p, 0,\ldots, 0)$, i.e., all the control parameters in $\vec{\Delta c_{1,2}}$ other than $\varphi_p$ are set to $0$.
Then, we can construct the robustness objective function $O_\mathrm{avg}$ from $O$ according to~\autoref{eq_robust_objective_by_average}.
We use the Adam optimizer~\cite{Kingma_adam} to minimize the objective functions $O$ and $O_{\mathrm{avg}}$. For the initial condition and optimizer hyperparameters listed in~\autoref{section_iswap_opt_details}, the objective functions during optimization are shown in~\autoref{fig_iswap_loss}, and the comparison of the result's robustness is shown in~\autoref{fig_opt_robust}.
While the optimization of $O_{\mathrm{avg}}$ has not converged, the result is good enough for the current experimental capabilities and demonstrating the viability of gradient optimization.
As shown in~\autoref{fig_opt_robust}, the parameters $\vec{p}_{\mathrm{avg}}$ obtained by optimizing $O_{\mathrm{avg}}$ are indeed more robust with respect to the errors in $\varphi_p$ compared to $\vec{p}$ obtained by optimizing $O$, and the minimum is lower.
The latter effect was unintended.
Retrospectively, due to the change of the objective function, $\vec{p}_{\mathrm{avg}}$ escaped the local minimum of $\vec{p}$.

With this example, we show that the joint gradient optimization is viable for a practical problem.
Because both  capcitively coupled fluxonium qubits and the control pulse~\autoref{eq_rectangular_pulse} have been realized in the experiment~\cite{aql2021fluxonium} and their behaviours coincide with the numerical simulation, we expect the gain from doing the optimization can transfer to the experiments as long as we have the correct decoherence model.
For the optimization of objective $O$, we reach a local minimum of both the device and control parameters in a single run of the gradient optimization.
To understand the benefits of using reverse-mode gradient computations in the optimization, we can compare the speed of the reverse-mode gradient computation with the computation of finite differences (see~\autoref{subsect_autodiff} for an explanation of finite differences).
This is because finite differences are widely used to estimate the gradients in the implementation of popular optimizers.
For example, this is how the Broyden--Fletcher--Goldfarb--Shanno (BFGS) algorithm~\cite{nocedal2006numerical,byrd_limited_1995} in SciPy~\cite{SciPy-NMeth} estimates the gradients when they are not provided by the user.
The speedup is about three times for this small optimization problem with seven device parameters and three control parameters (see~\autoref{section_iswap_opt_details} for the running time).
In~\autoref{subsect_benchmarking_diagonalization}, we use the task of diagonalizing fluxonium Hamiltonians to demonstrate that the speedup of the gradient optimization will grow when the system size and the number of parameters grow.


\textit{Discussion} 
In this work, we extend the optimal control framework to superconducting circuit parameters. We showcase that we can optimize over the device and control parameters jointly for a figure of merit reflecting a design goal through reverse-mode gradient computation. 
This suggests that there is almost no downside in terms of the efficiency to add device parameters to the optimization variables.

We can further extend the proposed framework to other stages of quantum processor design. The Hamiltonian representation with device parameters such as $E_C$, $E_J$, and $E_L$ we discussed in the paper is an intermediate representation when designing a quantum processor. This representation is derived from the underlying lumped-element circuit representation, which is again derived from the processor's two-dimensional (2D)/three-dimensional (3D) layout representation. There could be more freedom or design choices in the lumped-element circuit or layout representations. Moreover, the layout representation allows us to estimate substantial losses, such as the dielectric loss, based on the electric field energy distribution on the interfaces with contamination. However, electromagnetic simulations, which are used to evaluate the layout design, are very computationally expensive. This means that the gradient evaluation of the Hamiltonian parameters with respect to the layout parameters may be of substantial importance, as they will significantly reduce the number of iterations. We leave this for future study. 

Most optimal control use cases focus on optimizing control solutions to realize a given quantum operation. However, for NISQ (noisy intermediate-scale quantum) applications or quantum error correction, the overall performance depends on native gates available on the device, the compilation scheme used, and other device aspects like the readout and reset. 

In~\autoref{subsect_adiabatic_cphase}, we illustrate how we can simultaneously optimize the $ZZ$-interaction on--off ratios on different pairs of transmon qubits on a square lattice amenable to fault-tolerant quantum error correction architectures such as the surface code.
This can be viewed as a first step toward task-oriented optimization.

The authors would like to thank Dawei Ding and Fei Yan for their constructive comments and Yaoyun Shi for his encouragement and proof-reading of the manuscript.

\bibliographystyle{apsrev4-2}
\bibliography{bib}

\appendix

\section{Comparison of simultaneous optimization and alternating optimization}
\label{section_opt_separately_simultaneous}

\begin{figure}
    \includegraphics[width=0.9\linewidth]{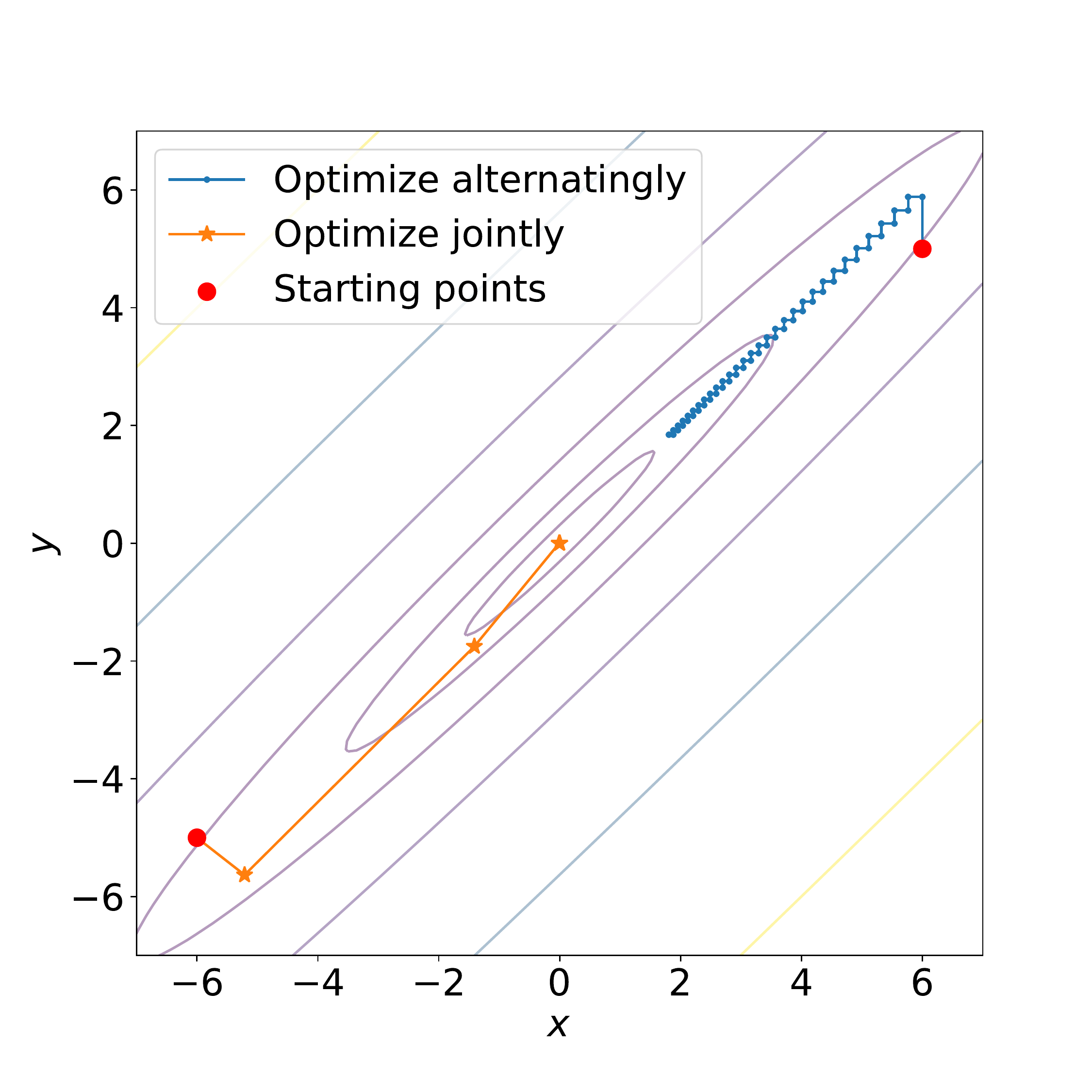}
    \caption{\label{fig_opt_sep_sim} Comparison of simultaneous optimization and optimization one at a time.
    The objective function is $f(x,y) = 100(x-y)^2+(x+y)^2$.
    The points corresponding to simultaneous optimization are from a single run of the Broyden--Fletcher--Goldfarb--Shanno (BFGS) optimization.
    The points corresponding to the separate optimizations are from a total number of 60 BFGS optimizations (30 for $x$ and 30 for $y$).
    }
\end{figure}

When using gradient optimization near a local minimum, it is generally better to optimize all the variables simultaneously instead of dividing them into groups and optimizing them one group at a time.
We can illustrate this point by considering a quadratic objective function $f(x)=v^T A v$, which provides a good approximation for general objective functions near their minima.
If we start the optimization with the coordinate $v_0$ such that $v_0$ is an eigenvector of $A$, then the natural descent direction is along the vector $-v_0$.
However, optimizing the elements of $v$ separately will force the optimization to move in less efficient directions.
~\autoref{fig_opt_sep_sim} shows a numerical simulation with the objective function $f(x,y) = 100(x-y)^2+(x+y)^2$, which confirms the above intuition.

It is not hard to construct toy gate optimization examples with similar objective function landscapes near local minima.
For example, we consider the problem of implementing a controlled-Z (CZ) gate using the mechanism in~\autoref{subsect_adiabatic_cphase}.
For simplicity, we assume that we can only use a (nearly) square pulse to tune the $gZZ$ interactions for time $t$, and the only penalty is decoherence.
The objective function can be selected as
\begin{equation}
    f(g,t) = (gt-\pi/2)^2 + t/T,
\end{equation}
where $T\gg 1$.
The descent direction is in the direction of decreasing $t$ while maintaining $gt\approx \pi/2$, which we cannot follow if we optimize them separately.

\section{Efficient gradient computation}
\label{subsect_autodiff}
To have a baseline to compare to, we examine the following naive method of computing gradients.
For a function $f(\vec{x})$, we can always the finite difference
\begin{equation}
    \left(f(\vec{x}+\epsilon \vec{e}_i)-f(\vec{x})\right)/\epsilon
\end{equation}
as a way to estimate the $i$-th element of the gradient, where $\vec{e}_i$ is the unit vector corresponding to the $i$-th direction.
The drawbacks of this method are as follows: 
\begin{enumerate}
    \item We need to evaluate the function $d+1$ times to compute the gradient, where $d$ is the dimension of $\vec{x}$.
    \item We need to have prior knowledge about $\max_x |f''(\vec{x})|$ for $\vec{x}$ in a certain neighborhood to know how to choose $\epsilon$.
    \item The smaller $\epsilon$ is, the higher the precision of $f$ needed to achieve the same level of precision in the final estimation becomes. In theory, this means we need longer computation times for small $\epsilon$. In practice, this can be more complicated because commonly used numerical computation packages such as NumPy do not provide an option for arbitrary precision.
\end{enumerate}

In this work, we use a more efficient and accurate method to compute the gradients.
This is usually called reverse-mode differentiation in the literature~\cite{giles_collected_2008}.
The method utilizes the decomposition of the function $f$,
\begin{equation}
    f = f_n \circ \cdots \circ f_1,
    \label{eq_general_function_decomposition}
\end{equation}
and then the chain rule can be used to compute the gradient of $f$.

If all the $f_i$'s are scalar valued, the reverse-mode differentiation can be performed in a single run, i.e., we need to compute the gradient of each $f_i$ once.
For the more general case of vector valued functions, let us consider the example
\begin{equation}
    \vec{x}\in \mathbb{R}^i \xrightarrow{f} \vec{y}\in\mathbb{R}^j \xrightarrow{g} \vec{z}\in \mathbb{R}^k .
\end{equation}
The partial derivatives of $f$ and $g$ form the Jacobian
\begin{align}
    J^{f}_{lm} &= \frac{\partial y_l}{\partial x_m}, \\
    J^{g}_{nl} &= \frac{\partial z_n}{\partial y_l}.
\end{align}
We know $\frac{\partial z_n}{\partial x_m} = \sum_l \frac{\partial z_n}{\partial y_l}\frac{\partial y_l}{\partial x_m}$.
In other words, $J^{g\circ f}_{nm} = J^g_{nl} J^f_{lm} $.
For most operations used in the examples of this paper, the computational complexities of $f_i$ and $f'_i$ are of the same order.
Theoretically, this is related to the fact that in the function decomposition~\autoref{eq_general_function_decomposition}, we can require functions $\{f_i\}$ to represent elemental computation steps, e.g., addition and multiplication.
For this finite set of elemental steps, the costs of computing the derivatives are bounded by some constant times the costs of performing the steps.
The above argument can be viewed as a sketch of the Baur--Strassen theorem~\cite{baur_complexity_1983}.
Therefore, we can typically expect a $d$-fold speedup with reverse-mode differentiation compared to the finite difference approach.
In~\autoref{subsect_benchmarking_diagonalization}, we perform some benchmarking for a common task, which also confirms the above conclusion.

Many matrix derivative formulas can be found in a previous publication~\cite{giles_collected_2008}.
For demonstration, we consider the example of computing derivatives for matrix diagonalization, which is a key step for the computing spectral properties of Hamiltonians.
Suppose that we have performed the forward computation $D = U^{-1} A U$, where $D$ is a diagonal matrix.
Then, $\mathrm{d}D = \operatorname{diag}(U^{-1} \mathrm{d}A \cdot U)$, where the $\operatorname{diag}(\cdot)$ operator set all the non-diagonal elements to $0$.
For $U$, we have $\mathrm{d}U = U(F\circ (U^{-1} \mathrm{d}A \cdot U))$, where $F_{ij} = (D_{jj}-D_{ii})^{-1}$ and $\circ$ is the operator for element-wise multiplication.
These derivatives are essentially the first-order perturbation theory.
The Jacobian can be extracted from these expressions.

We also want to discuss the compatibility of the differentiation and the control flows in programming languages, as there are misconceptions that using control flows will make a program non-differentiable.
We consider the popular activation function $\mathrm{ReLU}(x) = \max (x,0)$ that appeared in~\autoref{eq_rectangular_pulse}.
The derivative is
\begin{equation}
    \operatorname{ReLU}'(x)= 
\begin{cases}
    1,& \mathrm{if} \  x > 0\\
    0,              & \mathrm{if} \  x < 0
\end{cases}
\end{equation}
The derivative does not exist at $x=0$, but this fact does not stop a reverse-mode differentiation algorithm from converting ReLU to $\operatorname{ReLU}'$ by defining $\operatorname{ReLU}'(0)$ to be some arbitrary number.
A typical numerical optimization process is unlikely to reach the non-differentiable point $x=0$, so it may not matter in practice how we define $\operatorname{ReLU}' (0)$.
The same argument also holds for computing derivatives of the matrix diagonalization when the matrix has degenerate eigenvalues. For this work, we used the Python library JAX~\cite{jax2018github} to perform the reverse-mode differentiation.

\section{Benchmark of exact diagonalization}
\label{subsect_benchmarking_diagonalization}

In this subsection, we use the task of exact diagonalization to demonstrate the scaling of the speedup as the number of fluxonium qubits increases.
Consider the Hamiltonian
\begin{equation}
    H = \sum_i H_{f,i} + \sum_{\<i, i+1\>} J_{C,i} n_i n_{i+1},
\end{equation}
i.e., a 1-dimensional chain of capacitively coupled fluxonium qubits.
We construct the Hamiltonian by first truncating each fluxonium qubit to its lowest five (four for the case of six fluxonium qubits) energy levels.
We want to compute the gradient of the lowest eigenvalue w.r.t the parameters of each fluxonium and their coupling.
In~\autoref{table_ratio_time_gradient}, we list the ratios of the time needed for computing both the value and gradient to the time needed for only computing value.
We can see it is roughly a constant when we increase the number of qubits.
However, when we use finite difference to estimate the gradient, we need to compute the value $N$ times, where $N$ is the number of total parameters (see~\autoref{subsect_autodiff}).
Therefore, the speedup is roughly proportional to the number of qubits in the chain.
For the case of six fluxonium qubits, we will have an approximately $20\times$ speedup.

\begin{table}[h]
    \centering
    \begin{tabular}{|c|c|c|c|c|}
        \hline
        $n_{\mathrm{fm}}$ & 3 & 4 & 5 & 6 \\
        \hline
         ratio & 2.77 & 3.06 & 2.39 & 2.37\\ 
        \hline
    \end{tabular}
    \caption{Ratios of the time needed for computing both the value and gradient to the time needed for only computing value. The ratio is roughly a constant when we increase the number of fluxonium qubits $n_{\mathrm{fm}}$.}
    \label{table_ratio_time_gradient}
\end{table}

\section{Fidelity computation with phase compensation}
\label{subsection_fidelity_phase_compensation}
For multiple superconducting qubits, when we implement gates by tuning the flux, we inevitably change the energy difference between $\ket{0}$ and $\ket{1}$.
This will result in single qubit phases that need to compensated.
For example, we can perform the compensation through virtual $Z$ gates~\cite{mckay_efficient_2017} or real $Z$ gates~\cite{negirneac_high-fidelity_2021}.
For the case of using virtual $Z$, the fidelity is usually much higher than other operations, and thus, we can treat it as perfect.
Naturally, we want to compute the fidelity after single qubit phase compensation is done.
One way to perform the compensation, albeit not necessarily the best way, is as follows.
Assume the unitary operator we obtained from simulating the time evolution is $U$.
We set
\begin{equation}
    \phi_{ij} = \operatorname{arg}(U_{ij}),
\end{equation}
where $\operatorname{arg}(\cdot)$ computes the phase of a complex number.
We then compensate $U$ with a diagonal matrix $D$ equal to the tensor product of two single qubit phase gates:
\begin{equation}
    D = \operatorname{diag}\left(e^{-i\phi_{00}}, ie^{-i\phi_{21}},ie^{-i\phi_{12}}, -e^{-i(\phi_{12}+\phi_{21}-\phi_{00})}\right).
\end{equation}
Then, we can compute the average gate fidelity:
\begin{equation}
    F_\mathrm{avg}(UD,U_\mathrm{iSWAP}) = \frac{1}{d(d+1)} |\operatorname{tr} (UDU_\mathrm{iSWAP}^\dagger)+d|.
\end{equation}

\section{iSwap optimization details}
\label{section_iswap_opt_details}
In this section, we provide more details on the iSWAP gate optimization.
First, we examine the objective function
\begin{equation}
    O = \ln (1-F_m(U)+P_{\mathrm{decoh}}+P_\mathrm{fDiff}+P_\mathrm{fm}).
\end{equation}
For $P_{\mathrm{decoh}}$, we will consider two sources of decoherence mentioned previously ~\cite{nguyen2020toward}.
The first is the dielectric loss. By Fermi's golden rule, we can obtain the corresponding relaxation rate as
\begin{equation}
    \Gamma_d = \frac{1}{4E_C}| \bra{0} \hat{\varphi} \ket{1}|^2 \hbar\omega^2 \tan \delta_C \coth \left( \frac{\hbar\omega}{2k_B T} \right),
\end{equation}
where $\omega$ is the qubit angle frequency.
We set the effective dielectric loss tangent $\tan\delta_C= 2 \times 10^{-6}$ and the device temperature $T=50\ \textrm{mK}$.
The second is white flux noise, with the corresponding dephasing rate expressed as
\begin{equation}
    \Gamma_f = c_{f} \left( \frac{\partial \omega} {\partial \varphi_{\mathrm{ext}}} \right)^2,
\end{equation}
where we set $c_f = 3.95\times 10^{-15} \mathrm{s}$.
Here, we will not pursue a precise and accurate decoherence model, as it can vary drastically for different experimental setups.
Rather, all the losses listed in~\cite{nguyen_high-coherence_2019} have closed-form expressions that allow us to easily compute their gradients.
Together, we will use a penalty that is linear in time to approximate the infidelity caused by these two decoherence sources:
\begin{equation}
    P_{\mathrm{decoh}}=T_{\mathrm{gate}} (\Gamma_d / 2 + \Gamma_f).
\end{equation}
To obtain functioning single qubit gates, we consider a penalty $P_\mathrm{fDiff}$ that stops the frequency difference between the two qubits from being too small:
\begin{equation}
    P_\mathrm{fDiff} = c_\mathrm{fDiff} \operatorname{ReLU} (\Delta-|E_{01,1}-E_{01,2}|),
\end{equation}
where we set $\Delta = 0.1\ \mathrm{GHz}$, and $c_\mathrm{fDiff}=1\ \mathrm{GHz}^{-1}$.
The last penalty is $P_\mathrm{fm}$.
Considering the design feasibility, fabrication stability, and noise sensitivity, we set bounds for the variations of the fluxonium qubit parameters, as listed in~\autoref{table_iswap_parameter_bounds}.
\begin{table}[h]
    \centering
    \begin{tabular}{|c|c|c|}
        
        \hline
        Parameter & Lower bounds (GHz) & Upper bounds (GHz) \\
        \hline
        $E_{C}$   &  0.5 &  2.0       \\
        $E_{J}$   &  2.0   &  $+\infty$    \\
        $E_{L}$   &  0.5  &  1.5       \\
        \hline
    \end{tabular}
    \caption{Bounds for device parameters}
    \label{table_iswap_parameter_bounds}
\end{table}
For the optimization performed in this work, only the bound for $E_J$ will be violated.
Thus, we set 
\begin{equation}
    P_\mathrm{fm}=c_{\mathrm{fm},1}\sum_i \relu^2 (c_{\mathrm{fm},2} - E_{J,i}),
\end{equation}
where $c_{\mathrm{fm},1}=0.2\ \mathrm{GHz}^{-2}$ and $c_{\mathrm{fm},2} = 2.1\ \mathrm{GHz}$.

Here, we list the hyperparameters used in the Adam optimizer.
We set the initial learning rate to be $r_{\mathrm{init}} = 0.003$ for optimizing objective $O$ and $r_{\mathrm{init}} = 0.004$ for optimizing $O_{\mathrm{avg}}$.
In addition, we let the learning rate $r$ decay exponentially according to
\begin{equation}
    r = r_{\mathrm{init}} 2^{-\mathrm{step}/5000}.
\end{equation}
The reasoning is to increase the rate of the $O_{\mathrm{avg}}$ optimization and simultaneously reduce the fluctuations in the later steps.
Other hyperparameters for the Adam optimizer are $b_1=0.9$, $b_2=0.999$, and $\epsilon = 10^{-8}$.
There are $4\times 10^4$ gradient update steps for optimizing $O$ and $10^5$ steps for $O_{\mathrm{avg}}$.

The values of the parameters before and after optimization are listed in~\autoref{table_iswap_initial_values} and~\autoref{table_iswap_final_values}, respectively.
The initial values of the device parameters are chosen arbitrarily.
For the control parameters, we compute $\varphi_p$ such that it will ensure that the two fluxonium qubits are resonant.
It is chosen this way because if it is too far from resonance, the optimization landscape might be a barren plateau.
Determining the best way to perform the optimization without this kind of prior knowledge is an interesting subject, and we will examine it in future work.
\begin{table}[h]
    \centering
    \begin{tabular}{|c|c||c|c|}
        \hline
        \multicolumn{2}{|c||}{Device parameters} & \multicolumn{2}{|c|}{Control parameters}\\
        \hline
        Parameter & Value (GHz) & Parameter & Value \\
        \hline
        $E_{C,1}$   & 1.483 &  $t_\mathrm{ramp}$ & 2.31 ns \\ 
        $E_{J,1}$   & 2.082 &  $t_{\mathrm{plateau}}$   &  32.00 ns  \\ 
        $E_{L,1}$   & 0.626 &  $\varphi_p$   &  0.254   \\ 
        $E_{C,2}$   & 1.381 &     &                  \\ 
        $E_{J,2}$   & 2.103 &     &        \\ 
        $E_{L,2}$   & 0.620 &     &          \\ 
        $J_C $      & 0.143     &     &        \\ 

        \hline
    \end{tabular}
    \caption{Initial values of the parameters before optimization. The device parameters are from~\autoref{eq_time_dependent_capacitive_coupling}, and the control parameters are from~\autoref{eq_rectangular_pulse}. }
    \label{table_iswap_initial_values}
\end{table}
\begin{table}[h]
    \centering
    \begin{tabular}{|c|c||c|c|}
        \hline
        \multicolumn{2}{|c||}{Device parameters} & \multicolumn{2}{|c|}{Control parameters}\\
        \hline
        Parameter & Value (GHz) & Parameter & Value \\
        \hline
        $E_{C,1}$   & 1.492 &  $t_\mathrm{ramp}$ & 2.21 ns \\ 
        $E_{J,1}$   & 2.074 &  $t_{\mathrm{plateau}}$   &  31.14 ns  \\ 
        $E_{L,1}$   & 0.634 &  $\varphi_p$   &  0.245   \\ 
        $E_{C,2}$   & 1.412 &     &                  \\ 
        $E_{J,2}$   & 2.050 &     &        \\ 
        $E_{L,2}$   & 0.612 &     &          \\ 
        $J_C $      & 0.139     &     &        \\ 
        \hline
    \end{tabular}
    \caption{Normal iSwap optimization}
    \label{table_iswap_final_values}
\end{table}
\begin{table}[h]
    \centering
    \begin{tabular}{|c|c||c|c|}
        \hline
        \multicolumn{2}{|c||}{Device parameters} & \multicolumn{2}{|c|}{Control parameters}\\
        \hline
        Parameter & Value (GHz) & Parameter & Value \\
        \hline
        $E_{C,1}$   & 1.604 &  $t_\mathrm{ramp}$ & 1.88 ns \\ 
        $E_{J,1}$   & 2.101 &  $t_{\mathrm{plateau}}$   &  20.66 ns  \\ 
        $E_{L,1}$   & 0.714 &  $\varphi_p$   &  0.293   \\ 
        $E_{C,2}$   & 1.870 &     &                  \\ 
        $E_{J,2}$   & 2.095 &     &        \\ 
        $E_{L,2}$   & 0.532 &     &          \\ 
        $J_C $      & 0.207     &     &        \\ 
        \hline
    \end{tabular}
    \caption{Robustness iSwap optimization}
    \label{table_iswap_robust_values}
\end{table}

The computational times needed to compute the objective function and the gradient are listed in~\autoref{table_iswap_time_consumption}.
Since JAX provides just-in-time (\texttt{jit}) compilation functionality and it is applicable to our computation of $O$ and $\mathrm{\texttt{grad}} (O)$, we also list the computation times after using a \texttt{jit} compilation.
The speedup compared to the finite difference approach when not using \texttt{jit} is
\begin{equation*}
    T(\mathrm{loss} \  O) \cdot (n_{\mathrm{param}}+1) / T(\mathrm{\texttt{grad}} (O)) = 3.28,
\end{equation*}
while the speedup when using \texttt{jit} is
\begin{equation*}
    T(\mathrm{\texttt{jit}} (O)) \cdot (n_{\mathrm{param}}+1) / T(\mathrm{\texttt{jit}(\texttt{grad}} (O)) = 2.13.
\end{equation*}
\begin{table}[h]
    \centering
    \begin{tabular}{|c|c|}
        \hline
        Function & Running time (seconds) \\
        \hline
        loss $O$   &  $1.95\pm 0.07$  \\
        \texttt{grad}$(O)$   &  $6.53\pm 0.07$       \\
        \texttt{jit}$(O)$   &  $0.7 \pm 0.03$    \\
        \texttt{jit}(\texttt{grad}$(O)$)   &  $3.62\pm 0.06$       \\
        \hline
    \end{tabular}
    \caption{Elapsed real time for computing the loss function $O$ and its gradient. We also include the time needed for the \texttt{jit} compiled functions.}
    \label{table_iswap_time_consumption}
\end{table}
We see that we already obtain speedups with a very small number of parameters.
In~\autoref{subsect_benchmarking_diagonalization}, we will use the common task Hamiltonian diagonalization as a first attempt to benchmark the speedup of the reverse-mode gradient computation with respect to the number of parameters.

\section{Adiabatic flux-tuning gates}
\label{subsect_adiabatic_cphase}

As mentioned earlier, several popular cQED 2-qubit gate schemes are adiabatic~\cite{stehlik_tunable_2021,rol_fast_2019,yan_tunable_2018,sung_realization_2021}.
For these adiabatic gate schemes, we can obtain a good estimate of the gate performance solely based on the spectral properties of the Hamiltonian.
An example is that we can use the on--off ratio of the $ZZ$-interactions to benchmark the design of a tunable coupler.
More generally, there is a class of ``optimal design'' problems where we only optimize the design parameters.
Unsurprisingly, our gradient optimization techniques can also be applied to this class of problems, which we will demonstrate in the following example.
Moreover, we will use the example to show how our method can be applied to transmons and how it can be applied to a chip-level optimization problem.
The example is based on adiabatic flux-tuning CPhase gates obtained via $ZZ$-interactions caused by introducing one of the computational bases near an avoided crossing with a non-computational eigenstate.

For two capacitively coupled tunable transmons, the Hamiltonian has the form
\begin{equation}
    H(\varphi_{\mathrm{ext}}) = H_{t,1}(\varphi_{\mathrm{ext}}) + H_{t,2}(0) + J_C n_1 n_2\ ,
\end{equation}
where $H_{t,i}$ is the transmon Hamiltonian
\begin{equation}
    H_{t,i}(\varphi_{\mathrm{ext}}) = 4 E_{C,i} n_i^2 - \frac{1}{2}E_{J,i,\mathrm{eff}} (\varphi_{\mathrm{ext}})\sum_n (\ket{n}\bra{n+1} + \mathrm{h.c.}),
\end{equation}
with $E_{J,i,\mathrm{eff}}(\varphi_{\mathrm{ext}}) = E_{J,i} |\cos (\varphi_{\mathrm{ext}}/2)|$.
For simplicity, we have set all effective offset charges to be $n_g=0$.

The desired properties that we will try to optimize in this example are:
\begin{itemize}
    \item Large $ZZ$-interactions at the gate operating point;
    \item Sufficiently small $ZZ$-interactions at the idle point, so that single qubit gates including the identity gate have good fidelities;
    \item Long coherence times.
\end{itemize}
As we approach an avoided crossing between a computational basis and an eigenstate outside the computational subspace, the $ZZ$-interactions will be larger, and we need to change the Hamiltonian slower to avoid leakage.
For typical Hamiltonian parameters, we cannot reach the avoided crossing point adiabatically in a reasonable time compared to the coherence time.
Therefore, as a first attempt, we use the following routine to estimate the gate operating points:
\begin{enumerate}
    \item Initialize the flux $\varphi_{\mathrm{ext}} \leftarrow 0$ to the idle points.
    \item Changing the flux $\varphi_{\mathrm{ext}} \leftarrow \varphi_{\mathrm{ext}} + \Delta \varphi$.
    \item Use the following stop condition to check whether it is close to the avoided crossing. If not, go back to step 2.
\end{enumerate}
We choose the stop condition to be based on the overlap between the instantaneous eigenstates $ \{\psi_i(\varphi_{\mathrm{ext}})\} $ and the bare states $ \{\Psi_{jk}(\varphi_{\mathrm{ext}})\} $, where $1\leq i \leq 4$ and $1\leq j,k \leq 2$.
The instantaneous eigenstates $ \{\psi_i(\varphi_{\mathrm{ext}})\} $ are the eigenstates of the total Hamiltonian $H(\varphi_{\mathrm{ext}})$, and the bare states $ \{\Psi_{jk}(\varphi_{\mathrm{ext}})\} $ are the tensor products of $j$th and $k$th single transmon eigenstates.
The stop condition is then
\begin{equation}
    \min_i \sum_{jk} |\langle\psi_i(\varphi_{\mathrm{ext}})|\Psi_{jk}(\varphi_{\mathrm{ext}})\rangle|^2 < M,
    \label{eq_loop_stop_condition}
\end{equation}
where we set $M=0.8$ in this example.
Then, we can use the procedure specified in~\autoref{subsect_conditional_loop} to compute the gradient of the above algorithm with a conditional loop.

The magnitudes of the $ZZ$ interactions at the idle point and the gate operating point are computed from the energy of four computational basis states:
\begin{equation}
    E_{ZZ}(\varphi_{\mathrm{ext}}) = |E_{00} + E_{11} -E_{01}-E_{10}|,
    \label{eq_e_zz}
\end{equation}
where $E_{jk}$ are the eigenvalues corresponding to instantaneous eigenstates by $ \{\psi_i(\varphi_{\mathrm{ext}})\} $, which has the largest overlap with $ \{\Psi_{jk}(\varphi_{\mathrm{ext}})\} $.
Because we never reach the middle of the avoided crossing, the above $E_{jk}$'s can be defined unambiguously.
Again, we will include a few additional terms in the objective function to estimate the infidelity caused by decoherence and ensure the functionality of single qubit operations.
The objective function for minimization is
\begin{align}
    O = & E_{ZZ}(\varphi_{\mathrm{ext}})/P_{\mathrm{decoh}}\cdot c_{\mathrm{scale}}+P_{ZZ,\mathrm{idle}} \nonumber \\
        & + \sum_i (P_{\mathrm{tm},i} + P_{\mathrm{ah},i}) + P_\mathrm{fDiff},
\end{align}
where $P_{\mathrm{decoh}}$ is a penalty determined by the coherence time, $P_{ZZ,\mathrm{idle}}$ punishes large $ZZ$-interactions when qubits are at sweet spots, $P_{\mathrm{tm},i}$ ensures the $E_J / E_C$ ratio is large, $P_{\mathrm{ah},i}$ requires the anharmonicity to be large enough for single qubit gates, and $P_\mathrm{fDiff}$ forces the $0 \leftrightarrow 1$ frequencies of the two transmons to be different. The exact formulas for these terms can be found in~\autoref{section_adiabatic_opt_details}. In the above equation, $\varphi_{\mathrm{ext}}$ is determined by~\autoref{eq_loop_stop_condition}.

\begin{figure}
    \includegraphics[width=0.9\linewidth]{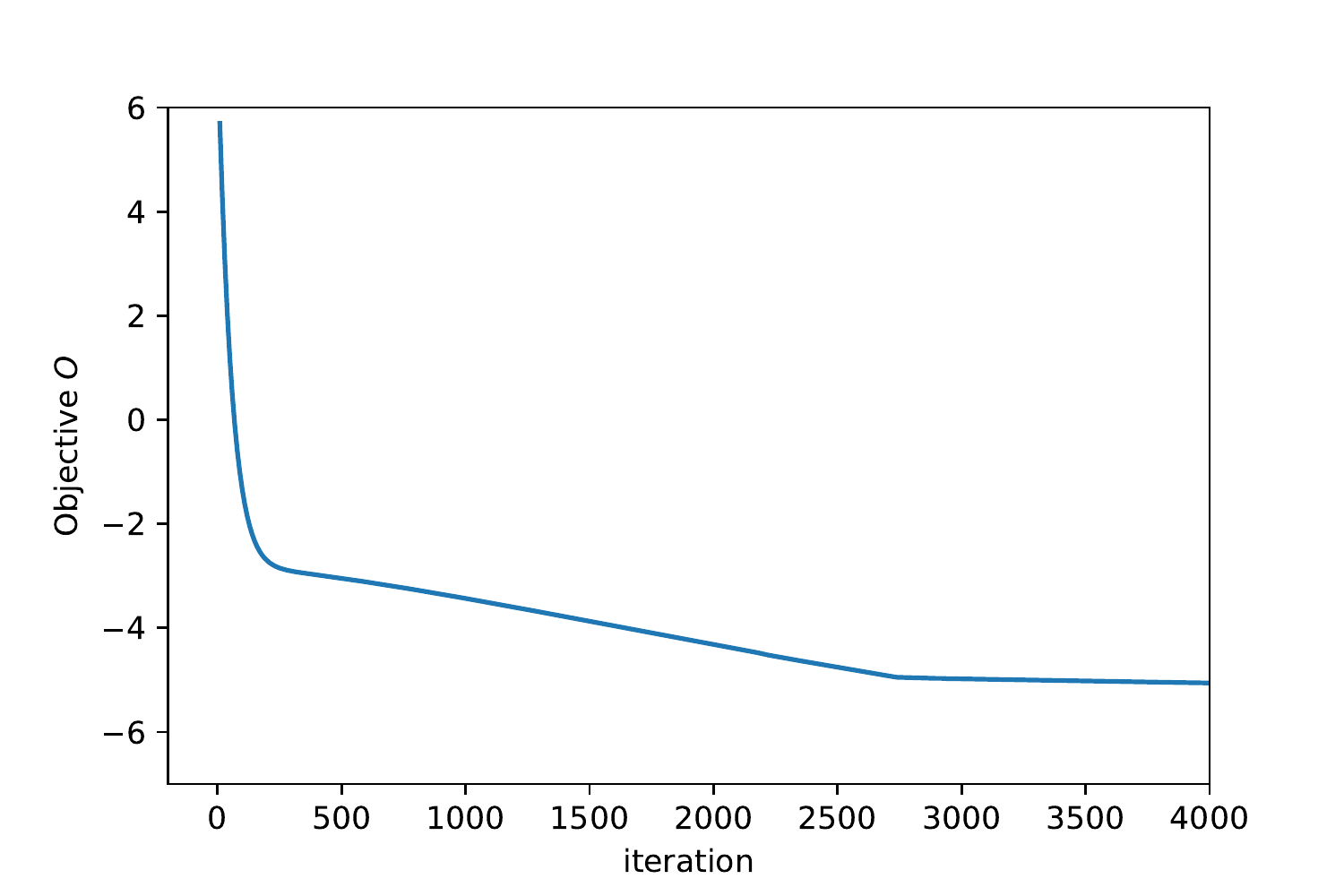}
    \caption{\label{fig_adiabatic_cphase_loss} Optimization of $O_{\mathrm{chip}}$ defined in~\autoref{eq_transmon_chip_obj}.}
\end{figure}

We can extend this optimization problem to a chip-level setting.
We consider the surface code scheme presented previously ~\cite{versluis_scalable_2017}.
In short, the plan is to have three types of transmon qubits with different parameters arranged in a certain way on a 2D lattice.
Sorting by their $0 \leftrightarrow 1$ transition frequencies, we label the three different types of transmon qubits as H(igh), M(iddle), and L(ow).
We want to have a CPhase gate between the H-M and M-L qubit pairs, while we still want the crosstalk to be small when idling.
Since H and L transmons are spatially separated, we will assume their interactions are small and ignore them in this example.
Therefore, we will add the objective function for H-M and H-L qubit pairs together:
\begin{equation}
    O_{\mathrm{chip}} = O_\mathrm{H-M} + O_\mathrm{M-L} - P_{\mathrm{tm, M}} - P_{\mathrm{ah, M}}.
    \label{eq_transmon_chip_obj}
\end{equation}
We need to subtract the last two terms because they are counted twice in $O_\mathrm{H-M} $ and $ O_\mathrm{M-L}$.
We can then use Adam~\cite{Kingma_adam} to perform the optimization.
The loss during optimization is shown in~\autoref{fig_adiabatic_cphase_loss}, and details about the optimization can be found  in~\autoref{section_adiabatic_opt_details}.

\section{Computing gradients for conditional loops}
\label{subsect_conditional_loop}

Here, we explain how to perform the differentiation in the presence of the conditional loop that is used in the adiabatic flux-tuning gate optimization.
The forward computation, i.e., the computation of the objective $O$, is carried out as usual.
When doing the reverse computation for the gradient, we utilize the fact that the gate operating point $\varphi^\mathrm{stop}_{\mathrm{ext}}$ is a function of parameters $\vec{p}$ in the optimization.
The equation
\begin{equation}
    \min_i \sum_{jk} |\langle\psi_i(\varphi^\mathrm{stop}_{\mathrm{ext}},\vec{p})|\Psi_{jk}(\varphi^\mathrm{stop}_{\mathrm{ext}}, \vec{p})\rangle|^2 = M
\end{equation}
implicitly defines the function $\varphi^\mathrm{stop}_{\mathrm{ext}}(\vec{p})$.
We can then use the implicit function theorem to compute the gradient of $\varphi^\mathrm{stop}_{\mathrm{ext}}(\vec{p})$ with respect to $\vec{p}$.
In a previous publication ~\cite{blondel_efficient_2021}, how to integrate the implicit function theorem with auto-differentiation frameworks is explained in more detail.

\section{Adiabatic CPhase gate optimization details}
\label{section_adiabatic_opt_details}

In this section, we will give more details on the adiabatic CPhase gate optimization.
First, we consider the objective function
\begin{align}
    O = & E_{ZZ}(\varphi_{\mathrm{ext}})/P_{\mathrm{decoh}}\cdot c_{\mathrm{scale}}+P_{ZZ,\mathrm{idle}} \nonumber \\
        & + \sum_i (P_{\mathrm{tm},i} + P_{\mathrm{ah},i}) + P_\mathrm{fDiff}.
\end{align}
To achieve low $ZZ$-crosstalk when both qubits are idling at sweet spots, we introduce the penalty $P_{ZZ,\mathrm{idle}}$ in the above objective function as follows:
\begin{equation}
    P_{ZZ,\mathrm{idle}} = c_{zz,1} \relu ((E_{ZZ}/h- c_{zz,2}) ),
\end{equation}
where $c_{zz,1} = 8\times 10^{-6}\ \hz^{-1}$ and $c_{zz,2} = 10^5\ \hz$.
For the penalty $P_{\mathrm{decoh}}$ related to the decoherence time, we first assume that $T_1$ is inversely proportional to the $0-1$ transition energy $E_{01}$.
With this approximation, we set
\begin{equation}
    P_{\mathrm{decoh}} = (E_{01,1}+E_{01,2}).
\end{equation}
We also set $c_{\mathrm{scale}} = 3000$ to ensure that $O$ is in a normal scale for optimization.
In addition, to obtain a low charge dispersion, we consider a penalty of the form
\begin{equation}
    P_{\mathrm{tm}} = \relu (c_{\mathrm{tm}}-E_J/E_C),
\end{equation}
where $c_{\mathrm{tm}} = 50$.
To have a reasonably large anharmonicity for single qubit operation, we add a penalty
\begin{equation}
    P_{\mathrm{ah}} = c_{\mathrm{ah},1} \relu^2 (c_{\mathrm{ah},2}-E_C/h),
\end{equation}
where $c_{\mathrm{ah},1} = 300\ \hz^{-2}$ and $c_{\mathrm{ah},2} = 3\times 10^8\ \hz$.
Similar to the previous example, a penalty related to the difference of the frequency is included, as follows:
\begin{equation}
    P_\mathrm{fDiff} = c_{\mathrm{fDiff},1}\operatorname{ReLU}^2 (c_{\mathrm{fDiff},2}-|E_{01,1}-E_{01,2}|/h),
\end{equation}
where $c_{\mathrm{fDiff},1}=5\times10^{-16}\ \hz^{-2}$ and $c_{\mathrm{fDiff},2}=2\times 10^8\ \hz$.

For the Adam optimization, we set the initial learning rate to be $8\times 10^{-5}$.
There is no learning rate decay.
The other hyperparameters are the same as in~\autoref{section_iswap_opt_details}.

In~\autoref{table_cz_optimization_values}, we list the important parameters before and after optimization for the adiabatic CPhase gate.
We also include the computed $ZZ$ interaction magnitudes $E_{ZZ}$ defined in~\autoref{eq_e_zz}.
\begin{table}[h]
    \centering
    \begin{tabular}{|c|c|c|}
        \hline
        Parameter & Value before (GHz) &  Value after (GHz) \\
        \hline
        $E_{C,H}$   &  0.3 &  0.437 \\
        $E_{J,H}$   &  21.5   & 21.87  \\
        $E_{C,M}$   &  0.28 &   0.285              \\
        $E_{J,M}$   &  16.5     &   16.51    \\
        $E_{C,L}$   &  0.24 &     0.257            \\
        $E_{J,L}$   &  14.5     &   14.15    \\
        $E_{ZZ}$, idle, H-M & $8.71\times 10^{-4}$ & $3.70\times 10^{-4}$ \\
        $E_{ZZ}$, idle, M-L & $1.62\times 10^{-3}$ & $1.02\times 10^{-4}$ \\
        $E_{ZZ}$, gate, H-M & $1.94\times 10^{-2}$ & $2.14\times 10^{-2}$ \\
        $E_{ZZ}$, gate, M-L & $2.19\times 10^{-2}$ & $5.37\times 10^{-3}$ \\
        \hline
    \end{tabular}
    \caption{Values before and after optimization for adiabatic CPhase gate.}
    \label{table_cz_optimization_values}
\end{table}

\end{document}